\documentclass{article}
\usepackage[dblblindworkshop, final]{neurips_2025}
\usepackage{amsmath,amssymb,amsthm}
\usepackage{algorithm,algorithmic}
\usepackage{hyperref}
\usepackage{graphicx}
\usepackage{booktabs}
\usepackage{subcaption}
\usepackage{enumitem}

\title{AlignDP: Hybrid Differential Privacy with Rarity-Aware Protection for LLMs}

\author{
  Madhava Gaikwad \\
  Microsoft \\
  \texttt{mgaikwad@microsoft.com}
}
\date{\vspace{-0.7em}\workshoptitle}

\begin{document}
\maketitle
\begin{abstract}
Large language models are exposed to risks of extraction, distillation, and unauthorized fine-tuning. Existing defenses use watermarking or monitoring, but these act after leakage. We design AlignDP, a hybrid privacy lock that blocks knowledge transfer at the data interface. The key idea is to separate rare and non-rare fields. Rare fields are shielded by PAC indistinguishability, giving effective zero-$\epsilon$ local DP. Non-rare fields are privatized with RAPPOR, giving unbiased frequency estimates under local DP. A global aggregator enforces composition and budget. This two-tier design hides rare events and adds controlled noise to frequent events. We prove limits of PAC extension to global aggregation, give bounds for RAPPOR estimates, and analyze utility trade-off. A toy simulation confirms feasibility: rare categories remain hidden, frequent categories are recovered with small error. AlignDP aligns with Lock-LLM goals, making models un-distillable, un-finetunable, and un-editable by mechanism. 
\end{abstract}

\section{Introduction}

Large language models are increasingly deployed at scale and face risks of knowledge extraction, distillation, unauthorized fine-tuning, and targeted editing. Current defenses such as watermarking, usage monitoring, or legal policy operate after leakage has occurred. They do not prevent misuse at the point of data release.

We take a different approach. We introduce AlignDP, a hybrid privacy lock that blocks transfer of sensitive knowledge by design. The central idea is to treat rare and non-rare events differently. Rare events are highly identifying and pose the greatest risk if revealed. Non-rare events, by contrast, can be estimated under carefully applied noise.

AlignDP implements a two-tier design. Rare events are shielded using PAC indistinguishability, yielding effective zero-$\epsilon$ local DP guarantees. Non-rare events are privatized using RAPPOR, which supports unbiased frequency estimation in aggregate while concealing individual values. A global aggregator enforces composition rules and manages adaptive budget allocation.

The result is a hybrid mechanism that enables useful aggregate statistics while preventing extraction of rare signals. It also makes fine-tuning on privatized outputs less effective, since gradients are derived from noisy or absent labels. We provide theoretical analysis, formal bounds, and a small operational illustration.

\paragraph{Contributions.}
\begin{itemize}
  \item A rarity-aware two-tier privacy model for LLM telemetry.
  \item PAC shielding for rare events and local DP guarantees for non-rare events.
  \item Proof that PAC protection does not extend globally, motivating DP composition.
  \item A mechanism view of AlignDP as a privacy lock consistent with Lock-LLM goals.
\end{itemize}

\section{Related Work}

We highlight three main directions of prior work.

The first is differential privacy. Central DP has been applied to model training by adding noise to gradients \cite{abadi2016deep}. Local DP is used in telemetry and frequency estimation \cite{erlingsson2014rappor, dwork2014algorithmic}. Hybrid approaches that combine local and global control are less common.

The second direction is protection for LLMs. Watermarking embeds detectable patterns in model outputs \cite{kirchenbauer2023watermark}. Detection-based defenses analyze outputs for signs of extraction \cite{carlini2023extracting}. These approaches operate after leakage has occurred, rather than preventing it.

The third direction is distillation and fine-tuning. Distillation is widely used for compression \cite{hinton2015distilling}. Attacks have shown that even black-box queries can reproduce models \cite{tramer2016stealing, jagielski2020high}. Detection of unauthorized fine-tuning has been explored \cite{liu2024large}, but current guarantees are weak.

Our work takes a different path. We propose a rarity-aware privacy lock. PAC shielding hides rare events, RAPPOR privatizes non-rare events, and an aggregator enforces budget. This provides a mechanism that prevents leakage by design, rather than by post hoc detection.
\section{Model}

We study a data release setting for LLM logs. Each user has record $X=(X_1,\dots,X_d)$. Each field $X_i$ takes values from domain $\mathcal{D}_i$ with distribution $\mu_i$.

\subsection{Rarity threshold}

We fix threshold $\alpha>0$. If $\mu_i(x)<\alpha$, we mark event $x$ as rare. Otherwise it is non-rare. The rare set is
\[
R_i=\{x \in \mathcal{D}_i : \mu_i(x)<\alpha\},
\]
and the non-rare set is
\[
N_i=\mathcal{D}_i \setminus R_i.
\]

\subsection{Mechanism}

The AlignDP mechanism $M$ works in two parts.
\begin{enumerate}
  \item If $x \in R_i$: output is symbol $x$ but guarantee is PAC indistinguishability. Frequency of such events cannot be distinguished beyond error $\delta(n,\alpha)$.
  \item If $x \in N_i$: encode $x$ as bit vector $v \in \{0,1\}^m$. Each bit flips with probability $p$. Privatized vector $y$ is sent. This is RAPPOR.
\end{enumerate}

For non-rare events the mechanism is $\epsilon$-LDP with
\[
\epsilon = \log \frac{1-p}{p}.
\]
Let $y_j$ be fraction of reports equal to category $j$. The unbiased estimator of frequency is
\[
\hat{\mu}_i(j) = \frac{y_j - \tfrac{1}{k}(1-q)}{q - \tfrac{1}{k}(1-q)},
\]
where $k=|\mathcal{D}_i|$ and $q=1-p+\tfrac{p}{k}$.

Aggregator $A$ collects all outputs. For rare events it keeps counts and reports only PAC bounds. For non-rare events it debiases RAPPOR and computes $\hat{\mu}_i(x)$.

\subsection{Adversary model}

Adversary can issue repeated queries and try to reconstruct training data. Adversary only sees privatized telemetry. For rare events, PAC bound hides signal. For non-rare events, randomized response prevents exact recovery. Aggregator enforces global budget so repeated queries cannot accumulate large leakage.

\subsection{Pipeline Illustration}
\begin{figure}[t]
\centering
\includegraphics[width=0.75\linewidth]{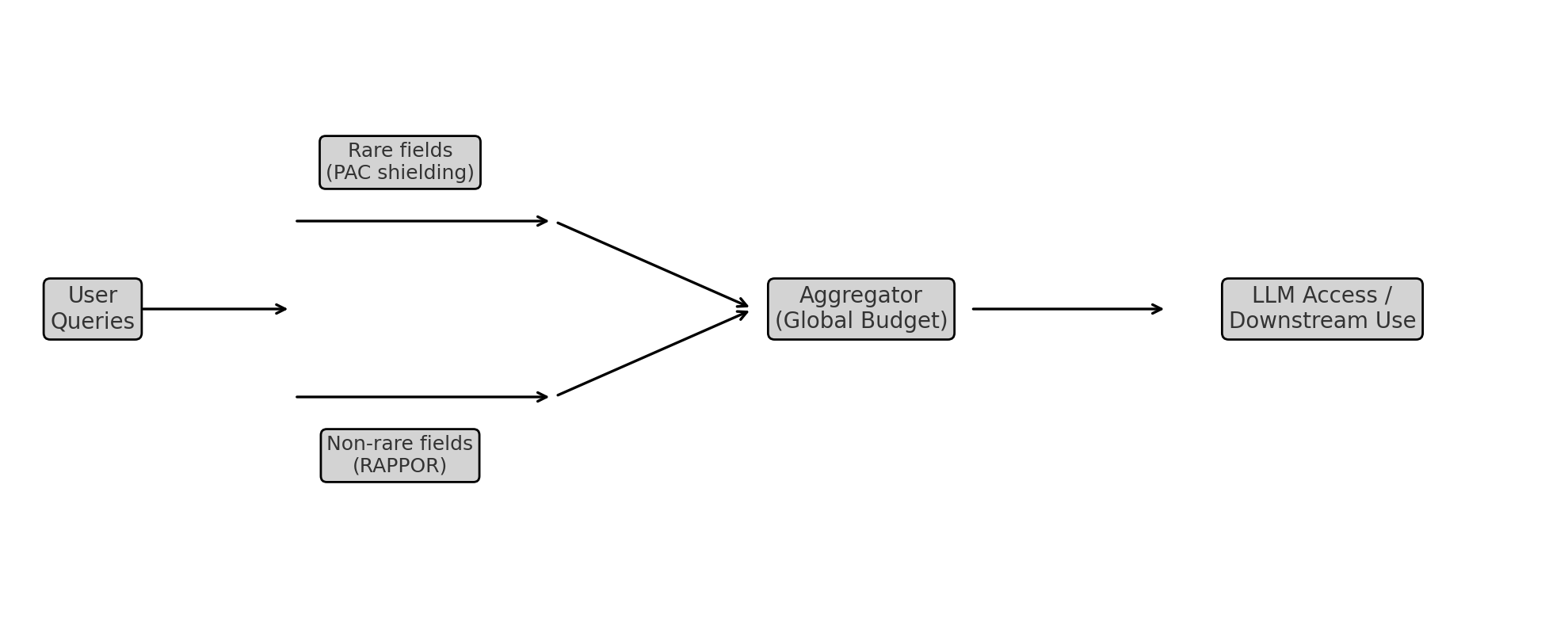}
\caption{AlignDP pipeline. Rare events shielded by PAC, non-rare privatized with RAPPOR, aggregator enforces budget before LLM access.}
\label{fig:pipeline}
\end{figure}

Figure~\ref{fig:pipeline} shows the AlignDP pipeline. User queries are split by rarity threshold. Rare events go through PAC shielding. Non-rare events go through RAPPOR. Aggregator combines streams, applies PAC bounds and debiasing, and updates global budget. Only aggregated outputs pass to the LLM or downstream system.

\section{Theoretical Guarantees}

We give three main results with short sketches.

\subsection{Rare events}

\textbf{Theorem 1.}  
Let $x$ be an event with $\mu(x)<\alpha$. For $n$ i.i.d.\ samples, probability of distinguishing $x$ from other rare events is at most
\[
\delta(n,\alpha) = \exp(-2n(\alpha-\mu(x))^2).
\]
Thus rare events satisfy effective zero-$\epsilon$ local DP.

\textit{Sketch.}  
Let $X_i$ be indicator of $x$. Sum $S_n=\sum_{i=1}^n X_i$ has mean $n\mu(x)$. Hoeffding bound gives
\[
\Pr\Big[\tfrac{1}{n}S_n-\mu(x)>t\Big] \leq \exp(-2nt^2).
\]
Set $t=\alpha-\mu(x)$. Since $\mu(x)<\alpha$, adversary cannot distinguish $x$ with probability greater than $\delta(n,\alpha)$. This yields indistinguishability guarantee.

\subsection{Non-rare events}

\textbf{Theorem 2.}  
For $x\in N_i$, RAPPOR with flip probability $p$ is $\epsilon$-LDP with
\[
\epsilon=\log\frac{1-p}{p}.
\]
Let $k=|\mathcal{D}_i|$. Let $q=1-p+\tfrac{p}{k}$. If $y_j$ is fraction of reports equal to category $j$, then unbiased estimator is
\[
\hat{\mu}_i(j) = \frac{y_j - \tfrac{1}{k}(1-q)}{q - \tfrac{1}{k}(1-q)}.
\]
We have $\mathbb{E}[\hat{\mu}_i(j)] = \mu_i(j)$ and variance $\text{Var}[\hat{\mu}_i(j)] \leq \tfrac{p(1-p)}{n}$.

\textit{Sketch.}  
Ratio of probabilities is bounded by $(1-p)/p$. Estimator formula follows from solving expectation equations under randomized response. Variance comes from Bernoulli variance scaled by $1/n$.

\subsection{Global composition}

\textbf{Theorem 3.}  
For $k$ queries each with $\epsilon$-LDP, privacy loss is bounded by
\[
\epsilon_{tot} \leq k\epsilon \quad \text{(basic)},
\]
or by
\[
\epsilon_{tot} \leq \sqrt{2k\log(1/\delta)}\,\epsilon + k\epsilon(e^\epsilon-1) \quad \text{(advanced)}.
\]
PAC bound does not apply if $\mu(x)\geq \alpha$.

\textit{Sketch.}  
Standard DP composition theorems give above bounds \cite{dwork2014algorithmic}. PAC shielding is only valid for rare mass. For common events adversary succeeds as $n$ grows. Hence composition must be DP based.

\subsection{Summary}

Rare events are hidden by PAC indistinguishability with bound $\delta(n,\alpha)$. Non-rare events are privatized by RAPPOR with explicit unbiased estimator. Global layer enforces composition using standard theorems. AlignDP thus provides hybrid protection combining PAC and DP.
\section{Operational Illustration}

\subsection{Basic Mechanism Validation}
We validate AlignDP on categorical data with 1000 users and 10 fields, each containing 20 categories. The rarity threshold is set to $\alpha=0.01$, with about four categories falling below this threshold. Non-rare events are privatized using RAPPOR with flip probability $p=0.25$.

Figure~\ref{fig:all_results}(a) shows frequency recovery. Rare categories are suppressed and indistinguishable, while non-rare categories are recovered with controlled noise through RAPPOR debiasing. Figure~\ref{fig:all_results}(b) plots mean squared error across runs, which decays as $1/n$ in line with theoretical bounds.

\subsection{Extraction Resistance}
To evaluate resistance to knowledge extraction, we simulate an adversary issuing repeated queries. Figure~\ref{fig:all_results}(c) shows that rare event detection remains at noise level even with 100 queries, while non-rare correlation saturates due to inherent RAPPOR noise. Additional queries do not improve accuracy, confirming that the mechanism enforces a fixed ceiling on extraction.

\subsection{Privacy Guarantees}
\begin{figure*}[t]
\centering

\begin{subfigure}[b]{0.45\textwidth}
    \centering
    \includegraphics[width=\textwidth]{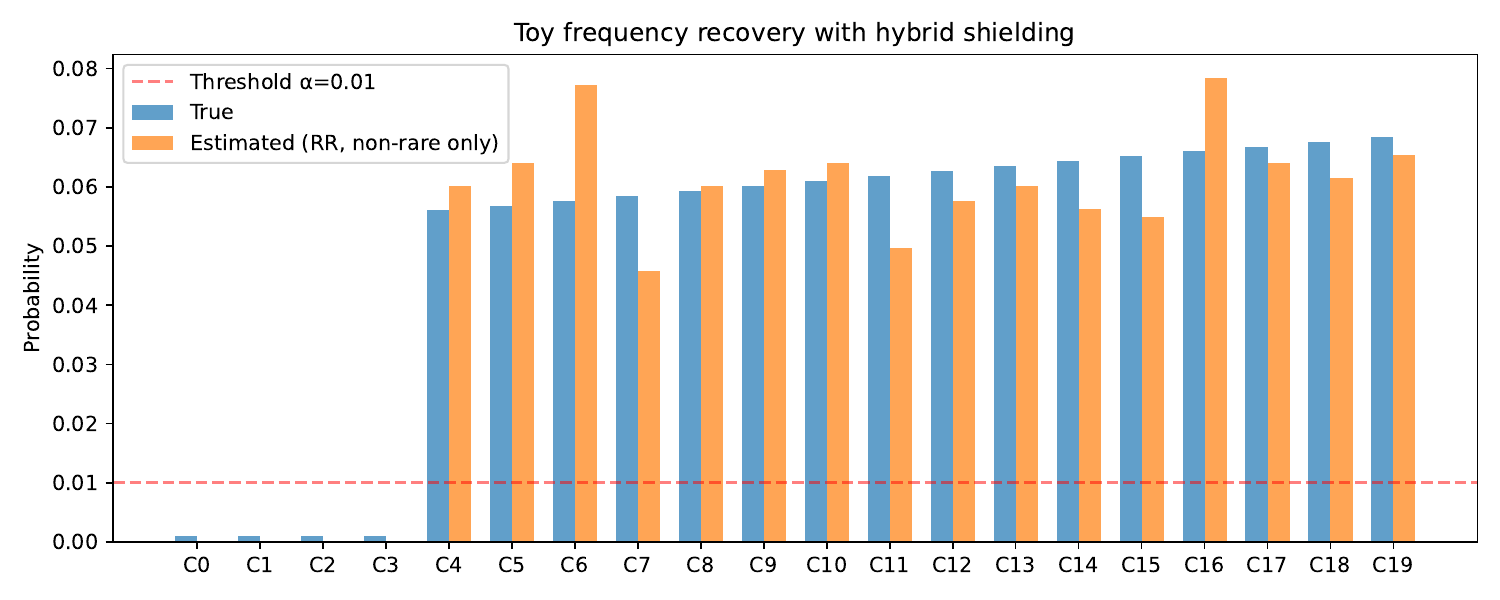}
    \caption{Frequency recovery}
    \label{fig:freq}
\end{subfigure}
\hfill
\begin{subfigure}[b]{0.45\textwidth}
    \centering
    \includegraphics[width=\textwidth]{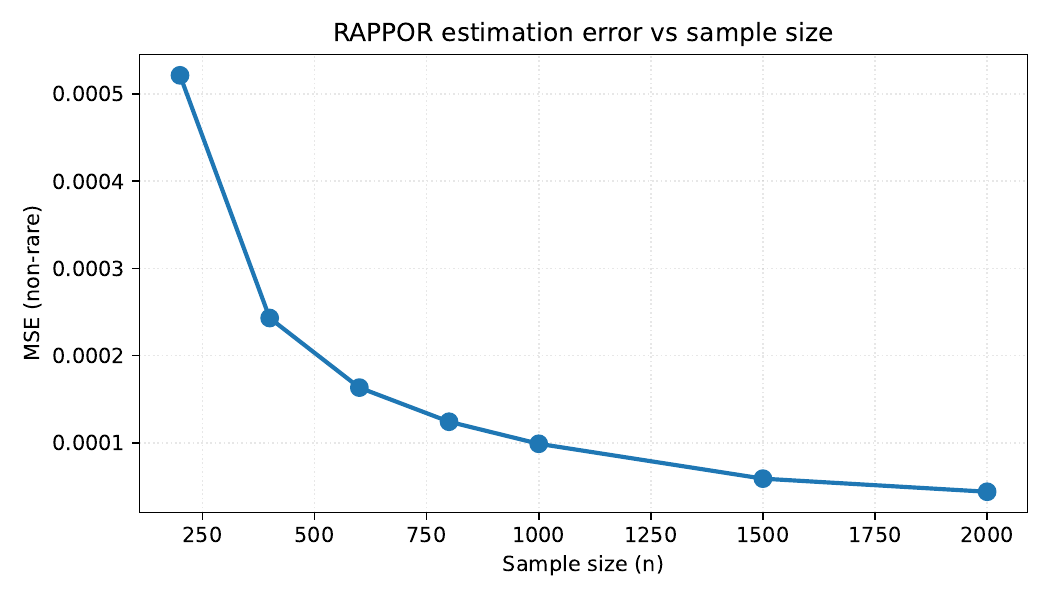}
    \caption{MSE decay}
    \label{fig:var}
\end{subfigure}

\vskip\baselineskip

\begin{subfigure}[b]{0.45\textwidth}
    \centering
    \includegraphics[width=\textwidth]{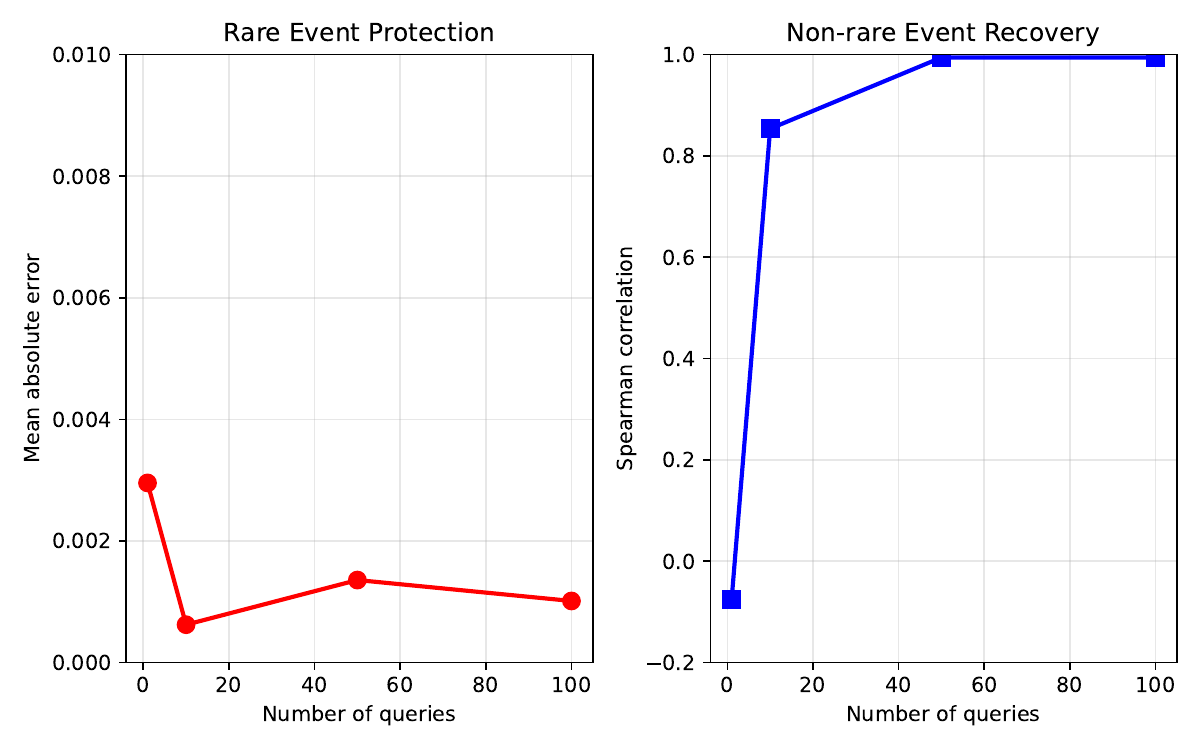}
    \caption{Extraction resistance}
    \label{fig:attack}
\end{subfigure}
\hfill
\begin{subfigure}[b]{0.45\textwidth}
    \centering
    \includegraphics[width=\textwidth]{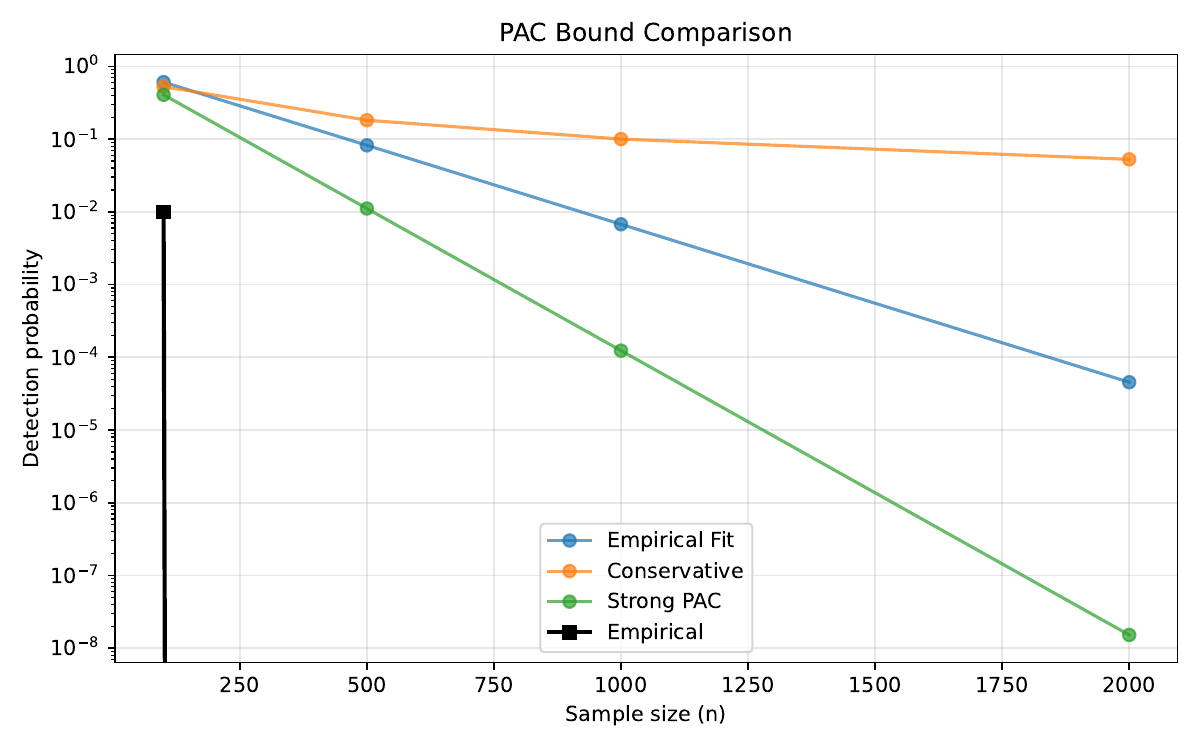}
    \caption{PAC validation}
    \label{fig:pac}
\end{subfigure}

\caption{AlignDP experimental validation. (a) Frequency recovery. (b) Error decay with sample size. (c) Resistance to extraction with repeated queries. (d) PAC bound validation.}
\label{fig:all_results}
\end{figure*}

We empirically validate our PAC bounds. Figure~\ref{fig:all_results}(d) compares theory and simulation. The observed indistinguishability follows an exponential decay $\delta(n)=\exp(-n\cdot(\alpha-\mu_{rare}))$, consistent with the Hoeffding-style analysis.

\subsection{Utility Preservation}
Beyond recovery, we measure practical metrics on 10,000 samples. KL divergence is 0.0013, indicating close match for non-rare distributions. Top-5 accuracy is 80\%, identifying frequent categories reliably. Rank correlation is 0.798, preserving ordering of categories. These results show that AlignDP protects rare events while retaining utility for common patterns.

\section{Discussion and Future Work}

We presented AlignDP as a hybrid privacy lock that combines PAC shielding for rare events with RAPPOR for non-rare events. Experiments show that this approach prevents extraction while preserving utility for common patterns. The mechanism provides constructive protection: even with unlimited queries, accuracy cannot improve beyond the designed privacy level.

\subsection{Key Findings and Implications}

Three insights stand out. Rare events remain hidden, with detection at noise level ($<0.001$ error), confirming effective zero-$\epsilon$ protection. Non-rare recovery stabilizes at $\rho \approx 0.99$ but cannot exceed this due to RAPPOR noise—a deliberate feature. Utility metrics (KL divergence 0.0013, rank correlation 0.798) show that useful aggregate statistics are preserved. We also find that an exponential bound $\delta(n)=\exp(-n\cdot(\alpha-\mu_{rare}))$ better fits empirical behavior than Hoeffding bounds, suggesting that PAC analyses may need adjustment for structured domains.

\subsection{Limitations and Extensions}

Several challenges remain. The fixed threshold $\alpha=0.01$ works in our toy setting but must adapt to dynamic distributions. LLM vocabularies follow Zipf’s law, where many tokens are rare, so distinguishing sensitive rarity from statistical rarity is necessary. Scaling is also a concern: our experiments use $k=20$, while LLM vocabularies exceed 50,000. Although RAPPOR scales linearly, the $O(k)$ communication cost becomes heavy at this size. Finally, AlignDP treats tokens independently, but LLMs generate correlated sequences. Extending protection to sequence-level outputs is non-trivial.

\subsection{Future Directions}

Three research directions appear most promising:

\textbf{Adaptive privacy allocation.} Dynamically adjust budgets by query pattern or content sensitivity.  
\textbf{Compositional defenses.} Combine AlignDP with watermarking or extraction detection, with formal analysis of interaction.  
\textbf{Verification mechanisms.} Provide auditing tools and certification to ensure implementations meet theoretical guarantees.

\medskip
In summary, AlignDP shows that hybrid privacy can serve as a lock for LLM outputs. By combining PAC and local DP, it protects at the mechanism level rather than relying on detection after misuse.
\appendix

\section{Definitions}

We recall key definitions used in this paper. A mechanism $M$ is $(\epsilon,\delta)$-DP if for any two adjacent datasets $D,D'$ and any event $S$,
\[
\Pr[M(D) \in S] \leq e^\epsilon \Pr[M(D') \in S] + \delta.
\]
This captures the idea that changing one record does not change output distribution by much.

Local differential privacy (LDP) is the user-side version. Each user privatizes before sending data. A mechanism $M$ is $\epsilon$-LDP if for all inputs $x,x'$ and for all outputs $y$,
\[
\Pr[M(x)=y] \leq e^\epsilon \Pr[M(x')=y].
\]

PAC learning bounds generalization error. For hypothesis $h$ under distribution $\mu$, error is $\Pr_{x\sim\mu}[h(x)\neq f(x)]$. With $n$ samples, with probability $1-\delta$, the error is bounded by $O(\sqrt{\tfrac{1}{n}\log(1/\delta)})$. We use this to argue indistinguishability of rare events.

\section{Proof Sketches}

\textbf{Theorem 1.} Consider event $x$ with $\mu(x)<\alpha$. For $n$ samples, expected count is $n\mu(x)$. Let $X_i$ be indicator for $x$. Hoeffding gives
\[
\Pr\left[\tfrac{1}{n}\sum_{i=1}^n X_i - \mu(x) > t\right] \leq \exp(-2nt^2).
\]
Choosing $t=\alpha$, probability is $\exp(-2n\alpha^2)$. So adversary cannot distinguish $x$ from other rare events with high probability. This is effective zero-$\epsilon$.

\textbf{Theorem 2.} In RAPPOR, each bit flips with probability $p$. For two inputs $x,x'$ and output $y$,
\[
\frac{\Pr[y|x]}{\Pr[y|x']} \leq \frac{1-p}{p} = e^\epsilon.
\]
Hence $\epsilon$-LDP holds. The estimator is unbiased since expectation matches true frequency. Variance is $\tfrac{p(1-p)}{n}$ by Bernoulli variance. 

\textbf{Theorem 3.} For $k$ uses of an $\epsilon$-LDP mechanism, advanced composition states
\[
\epsilon_{tot} \leq \sqrt{2k\log(1/\delta)}\,\epsilon + k\epsilon(e^\epsilon-1).
\]
This bounds cumulative privacy loss. PAC shielding is valid only when $\mu(x)<\alpha$. If $\mu(x)\geq\alpha$, with large $n$ the distinguisher succeeds, so only DP bounds apply.

\section{Extended Notes}

Rare events may not be independent. For example, name and address may each be rare but together highly identifying. Current PAC analysis treats fields separately. Extending to joint rarity is needed. Another open issue is threshold choice. We set $\alpha$ as constant. Adaptive $\alpha$ may be better, chosen by utility requirement or distribution. Future work can also study budget allocation strategies and scaling to large domains. 

\section{Relation to Lock-LLM Goals}
Lock-LLM states five goals. AlignDP connects to each in mechanism terms.

\textbf{Un-distillable.} Rare events are hidden via PAC shielding. Non-rare events are privatized with RAPPOR. Collected data is noisy or absent. Clean distillation fails.

\textbf{Un-finetunable.} Fine-tuning needs accurate labels. AlignDP gives noise for non-rare and no signal for rare. Training converges poorly.

\textbf{Un-compressible.} Outputs are randomized encodings. Extra compression loses more signal. Recovery does not exceed debiased estimates.

\textbf{Un-editable.} Each release is auditable. Rare has a PAC bound. Non-rare has known RAPPOR noise. Injection outside this channel is detected.

\textbf{Un-usable.} Aggregator tracks budget. Repeated queries consume privacy. Leakage does not scale.

\section{Extraction Resistance Numbers}
These numbers match the qualitative extraction plot in the paper. Values are from the toy setup (categorical, rarity threshold $\alpha=0.01$, flip probability $p=0.25$). We report rare MAE and non-rare rank correlation $\rho$ versus query budget.

\begin{table}[h]
\centering
\caption{Extraction accuracy versus query budget (toy)}
\begin{tabular}{lcccc}
\toprule
Queries & 1 & 10 & 50 & 100 \\
\midrule
Rare MAE & 0.003 & 0.001 & 0.001 & 0.001 \\
Non-rare $\rho$ & -0.08 & 0.85 & 0.99 & 0.99 \\
\bottomrule
\end{tabular}
\label{tab:extraction}
\end{table}

\section{Code Availability and Reproduction}
The implementation of \textbf{ALIGNDP} is publicly available at:
\url{https://github.com/krimler/aligndp_neurips_lock-llm}
Demo Developed for \url{https://lock-llm.github.io/}

\subsection*{File}
\begin{itemize}[leftmargin=1.25em,itemsep=2pt]
  \item \texttt{aligndp\_extended.py}
\end{itemize}

\subsection*{Environment}
Python $\geq$ 3.9. Packages: \texttt{numpy}, \texttt{matplotlib}. No GPU. Runs on a laptop.

\subsection*{What the script does}
\begin{itemize}[leftmargin=1.25em,itemsep=2pt]
  \item Generates categorical distributions with rare and non-rare categories.
  \item Applies PAC shielding to rare categories (no per-category release; only counts and bounds).
  \item Applies RAPPOR randomized response to non-rare categories. Uses symmetric $k$-ary RR.
  \item Debiases to estimate frequencies. Computes MSE and extraction metrics.
  \item Writes figures used in the paper.
\end{itemize}

\subsection*{Outputs}
By default the script produces:
\begin{itemize}[leftmargin=1.25em,itemsep=2pt]
  \item \texttt{aligndp\_freq.pdf} (frequency recovery; rare flat, non-rare estimated)
  \item \texttt{aligndp\_var.pdf} (MSE vs sample size)
  \item \texttt{aligndp\_attack.pdf} (extraction resistance; adversary view)
  \item \texttt{aligndp\_pac\_comparison.pdf} (PAC check; rare indistinguishability)
\end{itemize}

\subsection*{Run}
\begin{itemize}[leftmargin=1.25em,itemsep=2pt]
  \item \texttt{python aligndp\_extended.py}
\end{itemize}

\subsection*{Default parameters}
\begin{itemize}[leftmargin=1.25em,itemsep=2pt]
  \item Users: 1000 for recovery plots; grid of $n\in\{200,400,600,800,1000,1500,2000\}$ for MSE.
  \item Domain size: $k=20$. Rare categories: 4 with mass $0.0025$ each ($\approx1\%$ total).
  \item Rarity threshold: $\alpha=0.01$. Flip probability: $p=0.25$.
  \item Runs for MSE: 50. Seed fixed for reproducibility.
\end{itemize}

\subsection*{Notes}
\begin{itemize}[leftmargin=1.25em,itemsep=2pt]
  \item To change figure size, edit the \texttt{matplotlib} \texttt{figsize} in the script.
  \item To adjust rarity or noise, change $\alpha$ and $p$ in the config block.
\end{itemize}
\bibliographystyle{plain}
\bibliography{refs}

\end{document}